\documentclass[a4paper,twocolumn]{esapub2005}

\usepackage{graphicx}

\usepackage{times}
\usepackage[numbers]{natbib}

\title{The role of GRB 031203 in clarifying the astrophysical GRB scenario}

\author[1,2]{Remo Ruffini}
\author[1,2]{Maria Grazia Bernardini}
\author[1,2]{Carlo Luciano Bianco}
\author[1,2]{Letizia Caito}
\author[1,3]{Pascal Chardonnet}
\author[1,2]{Maria Giovanna Dainotti}
\author[1,4]{Federico Fraschetti}
\author[1,2]{Roberto Guida}
\author[1,2]{Gregory Vereshchagin}
\author[1,2]{She-Sheng Xue}
\affil[1]{ICRANet and ICRA, Piazzale della Repubblica 10, I-65122 Pescara, Italy.}
\affil[2]{Dipartimento di Fisica, Universit\`a di Roma ``La Sapienza'', Piazzale Aldo Moro 5, I-00185 Roma, Italy.}
\affil[3]{Universit\'e de Savoie, LAPTH - LAPP, BP 110, F-74941 Annecy-le-Vieux Cedex, France.}
\affil[4]{Centre CEA de Saclay (Essonne), Gif-sur-Yvette, 91191 cedex, France.}

\begin{document}


\maketitle

\begin{abstract}
The luminosity and the spectral distribution of the afterglow of GRB 031203 have been presented within our theoretical framework \citep{031203}, which envisages the GRB structure as composed by a proper-GRB, emitted at the transparency of an electron-positron plasma with suitable baryon loading, and an afterglow comprising the ``prompt emission'' as due to external shocks. In addition to the GRB emission, there appears to be a prolonged soft X-Ray emission lasting for $10^6$--$10^7$ seconds followed by an exponential decay. This additional source has been called by us URCA-3. It is urgent to establish if this component is related to the GRB or to the Supernova (SN). In this second case, there are two possibilities: either the interaction of the SN ejecta with the interstellar medium or, possibly, the cooling of a young neutron star formed in the SN 2003lw process. The analogies and the differences between this triptych GRB 031203 / SN 2003lw / URCA-3 and the corresponding ones GRB 980425 / SN 1998bw / URCA-1 and GRB 030329 / SN 2003dh / URCA-2, as well as GRB 060218 / SN 2006aj are discussed.
\end{abstract}

\section{Introduction}

The Gamma-Ray Burst (GRB) interpretational paradigms have been expressed in three Letters \citep{rlet1,rlet2,rlet3} and can be here summarized as follows:\\
{\bf 1) The Relative Space-Time Transformation (RSTT) paradigm \citep{rlet1}:} The first paradigm emphasizes the relevance of having the correct space-time parametrization of the source, which necessarily implies the knowledge of the equations of motion of the system and its entire worldline.\\
{\bf 2) The Interpretation of the Burst Structure (IBS) paradigm \citep{rlet2}:} The second paradigm emphasizes the structure of the most general GRB as composed of a proper GRB (P-GRB) and an extended afterglow, and emphasizes as well the common origin of the short and long GRBs as originating both from a gravitational collapse to a black hole.\\
{\bf 3) The GRB-Supernova Time Sequence (GSTS) paradigm \citep{rlet3}:} The third paradigm emphasizes on the concept of induced gravitational collapse, which was initially considered an effect of GRB triggering the late phase evolution of the progenitor star leading to the supernova (SN). We shall present an evolution of this paradigm.

In the next sections we summarize the structure of our model, we outline some progress made in recent years and we emphasize the relevance of the GRBs connected with SNe for testing these paradigms. In particular, in section \ref{991216} we speak about GRB 991216 as the first test ground of our model and in section \ref{050315} we discuss GRB 050315 which has given a confirmation of our predictions. We are then entitled to apply our theoretical framework to the analysis of the prompt emission of GRBs associated with SNe. We proceed first to GRB 980425 in section \ref{980425}; we go then to GRB 030329 in section \ref{030329}; finally, in section \ref{031203} we discuss GRB 031203 where we show a theoretical prediction of an hard-to-soft spectral evolution. We summarize the general results of these GRBs associated with SNe in Tab. \ref{tabella} and we make some general conclusions about the general structure of GRB, SN and URCA sources as well as some recent additions to our GSTS paradigm about the induced gravitational collapse.

\section{Our theoretical model}\label{991216}

\begin{figure}
\includegraphics[width=\hsize,clip]{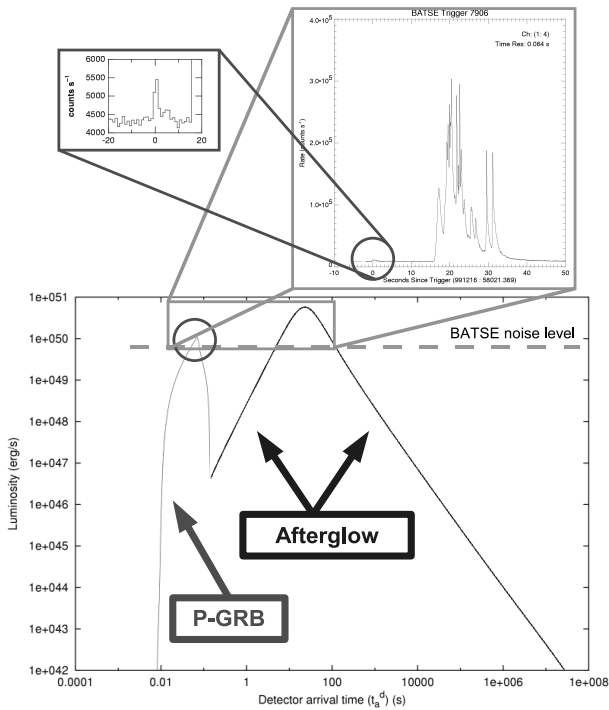}
\caption{GRB 991216 within our theoretical framework. The prompt emission observed by BATSE is identified with the peak of the afterglow, while the small precursor is identified with the P-GRB. Details in \citet{rlet2,rlet02,rubr,rubr2}.}
\label{991216_fig}
\end{figure}

Our GRB model, like all prevailing models in the existing literature \citep[see e.g.][and references therein]{p04,m02,m06}, is based on the acceleration of an optically thick electron-positron plasma. The specific issue of the origin and energetics of such a plasma, either in relation to black hole physics or to other physical processes, has often been discussed qualitatively in the GRB scientific literature but never quantitatively with explicit equations. The concept of the dyadosphere \citep{rukyoto,prx98} is the only attempt, as far as we know, to do this. This relates such an electron-positron plasma to black hole physics and to the characteristics of the GRB progenitor star, using explicit equations that satisfy the existing physical laws. Far from being just a theoretical discussion, this is essential to show that the physical origin and energetics of GRBs are the blackholic energy of the Kerr-Newman metric as defined by the \citet{cr71} mass-energy formula (see e.g. \citet{rubr2} and references therein, see also \citet{mtw73}).

Models of GRBs based on a single source (the ``collapsar'') generating both the SN and the GRB abounds in the literature \citep[see e.g.][]{wb06}. In our approach we have emphasized the concept of induced gravitational collapse, which occurs strictly in a binary system. The SN originates from a star evolved out of the main sequence and the GRB from the collapse to a black hole. The two phenomena are qualitatively very different. There is still much to be discovered about SNe due to their complexity, while the GRB is much better known since its collapse to a black hole is now understood. The concept of induced collapse implies at least two alternative scenarios. In the first, the GRB triggers a SN explosion in the very last phase of the thermonuclear evolution of a companion star \citep{rlet3}. In the second, the early phases of the SN induce gravitational collapse of a companion neutron star to a black hole \citep{mg11}. Of course, in absence of SN, there is also the possibility that the collapse to a black hole, generating the GRB, occurs in a single star system or in the final collapse of a binary neutron star system.

\begin{figure}
\includegraphics[width=\hsize,clip]{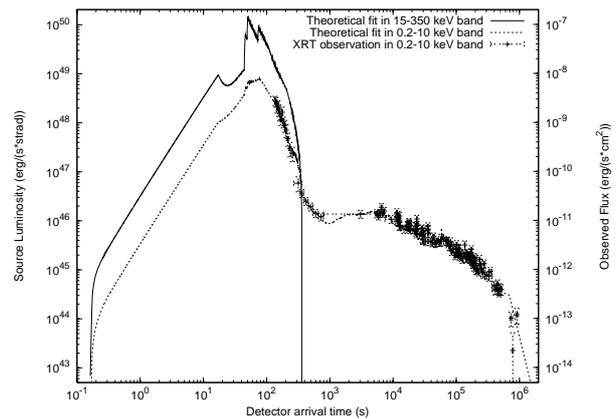}
\caption{Our theoretical fit (dotted line) of the XRT observations of GRB 050315 in the $0.2$--$10$ keV energy band \citep{va05}. The theoretical fit of the BAT observations in the $15$--$350$ keV energy band is also represented (solid line). See details in \citet{050315}.}
\label{050315_global}
\end{figure}

Turning now to the GRB origin, a major difference between our theoretical model and the ones in the current literature \citep[see e.g.][and references therein]{p04} is that what is usually called ``prompt emission'' in our case coincides with the peak of the afterglow emission and is not due to the prolonged activity of an ``inner engine'' which, clearly, introduces a jet undetermined physical process to explain the GRB phenomenon \citep{rlet2}. A basic feature of our model consists, in fact, in a sharp distinction between two different components in the GRB structure: {\bf 1)} the P-GRB, emitted at the moment of transparency of the self-accelerating $e^\pm$-baryons plasma \citep[see e.g.][]{g86,p86,sp90,psn93,mlr93,gw98,rswx99,rswx00,rlet1,rlet2,Monaco_RateEQ}; {\bf 2)} an afterglow described by external shocks and composed of three different regimes \citep[see][and references therein]{rswx99,rswx00,rlet2,rubr}. The first afterglow regime corresponds to a bolometric luminosity monotonically increasing with the photon detector arrival time, corresponding to a substantially constant Lorentz gamma factor of the accelerated baryons. The second regime consists of the bolometric luminosity peak, corresponding to the ``knee'' in the decreasing phase of the baryonic Lorentz gamma factor. The third regime corresponds to a bolometric luminosity decreasing with arrival time, corresponding to the late deceleration of the Lorentz gamma factor. In some sources the P-GRB is under the observability threshold. In \citet{rlet2} we have chosen as a prototype the source GRB 991216 which clearly shows the existence of the P-GRB and the three regimes of the afterglow (see Fig. \ref{991216_fig}). Unfortunately, data from BATSE existed only up to $ 36 $ s, and data from R-XTE and Chandra only after $ 3500 $ s, leaving our theoretical predictions in the whole range between $ 36 $ s and $ 3500 $ s without the support of the comparison with observational data. Nevertheless, both the relative intensity of the P-GRB to the peak of the afterglow in such source, as well as their corresponding temporal lag, were theoretically predicted within a few percent (see Fig. 11 in \citet{rubr}).

\section{GRB 050315}\label{050315}

Thanks to the data provided by the \emph{Swift} satellite, we were finally able to confirm, by direct confrontation with the observational data, our theoretical predictions on the GRB structure with a detailed fit of the complete afterglow light curve of GRB 050315, from the peak, including the ``prompt emission'', all the way to the latest phases without any gap in the observational data \citep[see][]{050315}. The best fit of the GRB 050315 observational data leads to a total energy of the black hole dyadosphere, generating the $e^\pm$ plasma, $E_{dya} = 1.46\times 10^{53}$ erg \citep[the observational \emph{Swift} $E_{iso}$ is $> 2.62\times 10^{52}$ erg, see][]{va05}, so that the plasma is created between the radii $r_1 = 5.88\times 10^6$ cm and $r_2 = 1.74 \times 10^8$ cm with an initial temperature $T = 2.05 MeV$ and a total number of pairs $N_{e^+e^-} = 7.93\times 10^{57}$. The second parameter of the theory, the amount $M_B$ of baryonic matter in the plasma, is found to be such that $B \equiv M_Bc^2/E_{dya} = 4.55 \times 10^{-3}$. The transparency point and the P-GRB emission occurs then with an initial Lorentz gamma factor of the accelerated baryons $\gamma_\circ = 217.81$ at a distance $r = 1.32 \times 10^{14}$ cm from the black hole.

\begin{figure}
\includegraphics[width=\hsize,clip]{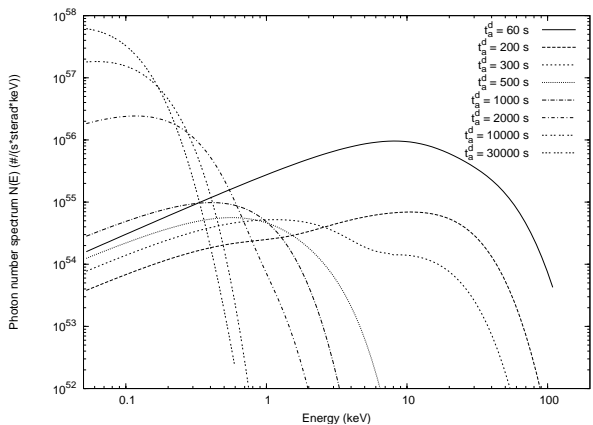}
\caption{Eight theoretically predicted instantaneous photon number spectra $N(E)$ are here represented for different values of the arrival time. The hard to soft behavior is confirmed. See details in \citet{050315}.}
\label{050315_spettro}
\end{figure}

It has been recently pointed out \citep{nousek} that almost all the GRBs observed by {\em Swift} show a ``canonical behavior'': an initial very steep decay followed by a shallow decay and finally a steeper decay. In order to explain these features many different approaches have been proposed \citep{meszaros,nousek,panaitescu,zhang}. In our treatment these behaviors are automatically described by the same mechanism responsible for the prompt emission, which is our case coincides with the peak of the afterglow. The baryonic shell expands in an interstellar medium (ISM) region from a radius $r = 4.15 \times 10^{16}$ cm up to $r = 5.62 \times 10^{16}$ cm with average effective ISM density $\langle n_{ism} \rangle = 0.81$ particle/$cm^3$. The remaining part of the afterglow appears to have a significantly lower effective ISM density, decreasing inversely proportional with the distance. We find the effective ISM density $\langle n_{ism} \rangle = 1.0\times 10^{-4}$ particle/$cm^3$ at $r = 5.67 \times 10^{16}$ cm and $\langle n_{ism} \rangle = 2.0\times 10^{-6}$ particle/$cm^3$ at $r = 5.98 \times 10^{18}$ cm. This produces a slower decrease of the velocity of the baryons with a consequent longer duration of the afterglow emission. The initial steep decay of the observed flux is due to the smaller number of collisions with the ISM. In Fig. \ref{050315_global} is represented our theoretical fit of the XRT data, together with the theoretically computed $15$--$350$ keV light curve \citep[details in][]{050315}. We attribute this decrease in the ISM effective density to a sudden fragmentation of the baryonic shell. What is impressive is that no different scenarios need to be advocated in order to explain the features of the light curves: both the prompt and the afterglow emission are just due to the inelastic collisions of the expanding baryonic shell with the ISM, which produce a radiation with a thermal spectrum in the co-moving frame \citep{spectr1}.

\begin{figure}
\includegraphics[width=\hsize,clip]{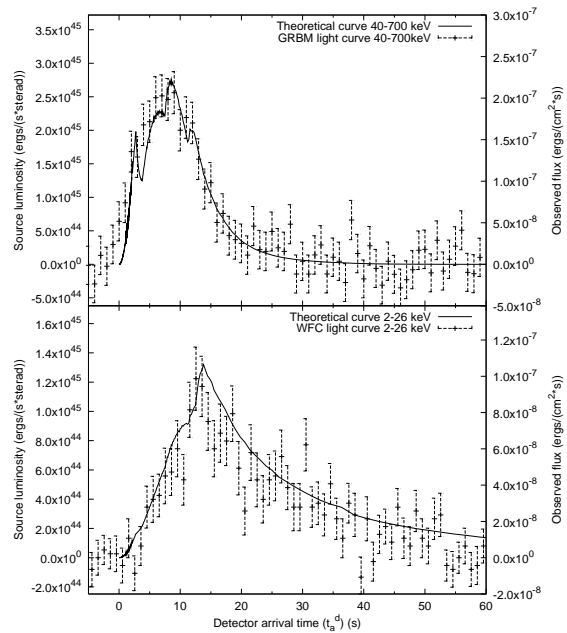}
\caption{Theoretical light curves of GRB 980425 prompt emission in the $40$--$700$ keV and $2$--$26$ keV energy bands (solid line), compared with the observed data respectively from Beppo-SAX GRBM and WFC \citep[see][]{pian00,frontera}.}
\label{980425_picco}
\end{figure}

In addition to the the luminosity in fixed energy bands we can derive also the instantaneous photon number spectrum $N(E)$. In Fig. \ref{050315_spettro} are shown samples of time-resolved spectra for eight different values of the arrival time which cover the whole duration of the event. It is manifest from this picture that, although the spectrum in the co-moving frame of the expanding pulse is thermal, the shape of the final spectrum in the laboratory frame is clearly non thermal. In fact, as explained in \citet{spectr1}, each single instantaneous spectrum is the result of an integration of millions of thermal spectra, each one duly weighted by appropriate Lorentz and Doppler factors, over the corresponding EQuiTemporal Surface \citep[EQTS, see][]{EQTS_ApJL,EQTS_ApJL2}. This calculation produces a non thermal instantaneous spectrum in the observer frame (see Fig. \ref{050315_spettro}).

A distinguishing feature of the GRBs spectra which is also present in these instantaneous spectra is the hard to soft transition during the evolution of the event \citep{cri97,frontera,gcg02}. In fact the peak of the energy distribution $E_p$ drifts monotonically to softer frequencies with time. This feature is linked to the change in the power-law low energy spectral index $\alpha$ \citep{b93}, so the correlation between $\alpha$ and $E_p$ \citep{cri97} is explicitly shown. The observed energy distribution changes from hard to soft, with continuity, from the ``prompt emission'' all the way to the latest phases of the afterglow.

\section{GRB 980425 / SN 1998bw / URCA-1}\label{980425}

\begin{figure}
\includegraphics[width=\hsize,clip]{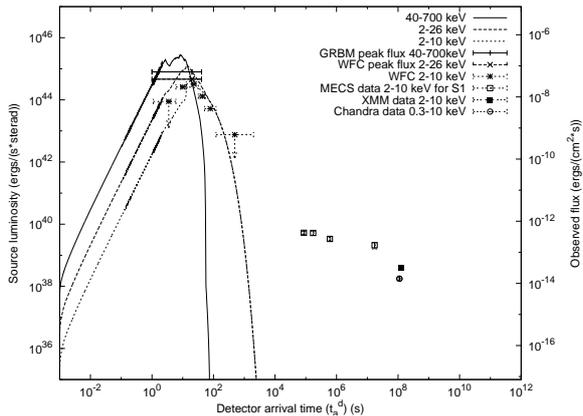}
\caption{Theoretical light curves of GRB 980425 in the $40$--$700$ keV (solid line), $2$--$26$ keV (dashed line), $2$--$10$ keV (dotted line) energy bands, represented together with URCA-1 observational data. All observations are by BeppoSAX \citep{pian00}, with the exception of the last two URCA-1 points, which is observed by XMM and \textit{Chandra} \citep{pian04,k04}.}
\label{980425_global}
\end{figure}

The best fit of the observational data of GRB 980425 \citep{pian00,frontera} leads to $E_{e^\pm}^{tot}=1.2\times10^{48}$ erg and $B = 7.7\times10^{-3}$. This implies an initial $e^\pm$ plasma with $N_{e^+e^-} = 3.6\times10^{53}$ and with an initial temperature $T = 1.2$ MeV. After the transparency point, the initial Lorentz gamma factor of the accelerated baryons is $\gamma = 124$. The variability of the luminosity, due to the inhomogeneities of the ISM \citep{rlet02}, is characterized by a density contrast $\delta n / n \sim 10^{-1}$ on a length scale of $\Delta \sim 10^{14}$ cm. We determine the effective ISM parameters to be: $\langle n_{ism} \rangle = 2.5\times 10^{-2}$ particle/$cm^3$ and $\langle \mathcal{R} \rangle = 1.2\times 10^{-8}$, where $\mathcal{R}\equiv A_{eff}/A_{vis}$ is the ratio between the effective emitting area $A_{eff}$ of the expanding shell and its entire visible area $A_{vis}$ and takes into account both the effective porosity of the shell and the ISM filamentary structure \citep{spectr1,fil}.

In Fig. \ref{980425_picco} we address the first $60$ s of data, which in the current literature are generally called ``prompt emission'' as due to an unidentified ``inner engine'' and neglected in the theoretical modeling of the source. In our approach, we test our specific theoretical assumptions comparing and contrasting our theoretically computed light curves in the $40$--$700$ and $2$--$26$ keV energy bands with the observations by the BeppoSAX GRBM and WFC during such time interval \citep[see][]{pian00,frontera}. As in the previous works, we have used our exact analytic solution for the equations of motion of the baryons \citep{PowerLaws}. The agreements in Fig. \ref{980425_picco} shows the very satisfactory predictive power of our theory.

\begin{figure}
\includegraphics[width=\hsize,clip]{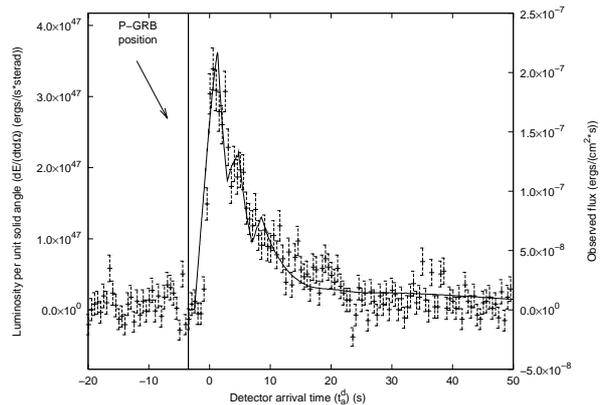}
\caption{Theoretically simulated light curve of the GRB 031203 prompt emission in the $20-200$ keV energy band (solid line) is compared with the observed data from \citet{saz}. The vertical bold line indicates the time position of P-GRB. See details in \citet{031203}.}
\label{031203_picco}
\end{figure}

In Fig. \ref{980425_global} we summarize some of the problematic implicit in the old pre-\emph{Swift} era: data are missing in the crucial time interval between $60$ s and $10^5$ s, when the BeppoSAX NFI starts to point the GRB 980425 location. In this region we have assumed, for the effective ISM parameters, constant values inferred by the last observational data. Currently we are relaxing this condition, also in view of the interesting paper by \citet{ga06}. In this respect, we are currently examining GRB 060218 / SN 2006aj \citep[see][]{caa06,060218}. We then represent the URCA-1 observations performed by BeppoSAX-NFI in the energy band $2$--$10$ keV \citep{pian00}, by XMM-EPIC in the band $0.2$--$10$ keV \citep{pian04} and by \textit{Chandra} in the band $0.3$--$10$ keV \citep{k04}. The separation between the light curves of GRB 980425 in the $2$--$700$ keV energy band, of SN 1998bw in the optical band \citep{nomoto,pian06}, and of the above mentioned URCA-1 observations is given in Fig. \ref{urca123+GRB_full}A. It remains to be determined by additional observations by Swift in other GRBs associated to SNe if, indeed, a connection between the prompt emission and the URCA sources can be established.

\section{GRB 030329 / SN 2003dh / URCA-2}\label{030329}

\begin{figure}
\includegraphics[width=\hsize,clip]{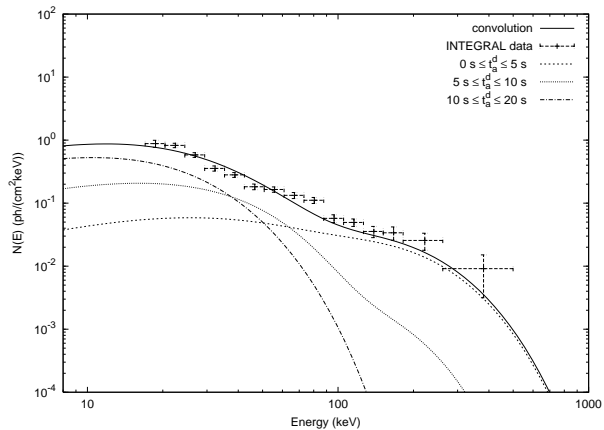}
\caption{Three theoretically predicted time-integrated photon number spectra $N(E)$ are here represented for $0 \le t_a^d \le 5$ s (dashed curve), $5 \le t_a^d \le 10$ s (dotted curve) and $10 \le t_a^d \le 20$ s (dash-dotted curve). The hard to soft behavior is confirmed. Moreover, the theoretically predicted time-integrated photon number spectrum $N(E)$ corresponding to the first $20$ s of the ``prompt emission'' (bold solid curve) is compared with the data observed by INTEGRAL \citep[see][]{saz,saz2}. This curve is obtained as a convolution of 108 instantaneous spectra, which are enough to get a good agreement with the observed data. See details in \citet{031203}.}
\label{031203_spettro}
\end{figure}

For GRB 030329 we have obtained a total energy $E_{e^\pm}^{tot}=2.12\times10^{52}$ erg and a baryon loading $B = 4.8\times10^{-3}$. This implies an initial $e^\pm$ plasma with $N_{e^+e^-}=1.1\times10^{57}$ and with an initial temperature $T=2.1$ MeV. After the transparency point, the initial Lorentz gamma factor of the accelerated baryons is $\gamma = 206$. The effective ISM parameters are $\langle n_{ism} \rangle = 2.0$ particle/$cm^3$ and $\langle \mathcal{R} \rangle = 2.8\times 10^{-9}$, with a density contrast $\delta n / n \sim 10$ on a length scale of $\Delta \sim 10^{14}$ cm. The resulting fit of the observations, both of the prompt phase and of the afterglow have been presented in \citep{030329,030329_la}. We compare in Fig. \ref{urca123+GRB_full}B the light curves of GRB 030329 in the $2$--$400$ keV energy band, of SN 2003dh in the optical band \citep{nomoto,pian06} and of URCA-2 observed by XMM-EPIC in $2$--$10$ keV energy band \citep{tiengo03,tiengo04}.

\section{GRB 031203 / SN 2003lw / URCA-3}\label{031203}

\begin{figure}[t]
\includegraphics[width=\hsize,clip]{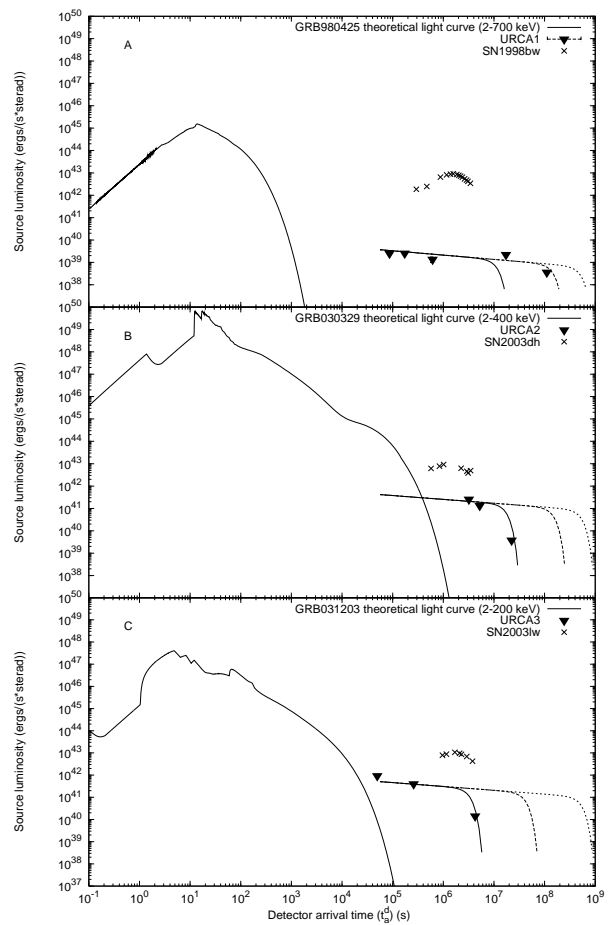}
\caption{Theoretically computed light curves of GRB 980425 in the $2$--$700$ keV band (A), of GRB 030329 in the $2$--$400$ keV band (B) and of GRB 031203 in the $2$--$200$ keV band (C) are represented, together with the URCA observational data and qualitative representative curves for their emission, fitted with a power law followed by an exponentially decaying part. The luminosity of the SNe in the $3000-24000$ {\AA} are also represented \citep{nomoto,pian06}.}
\label{urca123+GRB_full}
\end{figure}

The analysis of GRB 031203 leads to a total energy $E_{e^\pm}^{tot}=1.85\times10^{50}$ erg and to a baryon loading $B = 7.4\times10^{-3}$. This implies an initial $e^\pm$ plasma with $N_{e^+e^-}=3.0\times 10^{55}$ and with an initial temperature $T=1.5$ MeV. After the transparency point, the initial Lorentz gamma factor of the accelerated baryons is $\gamma = 132$. The effective ISM parameters are $\langle n_{ism} \rangle = 1.6\times 10^{-1}$ particle/$cm^3$ and $\langle \mathcal{R} \rangle = 3.7\times 10^{-9}$, with a density contrast $\delta n / n \sim 10$ on a length scale of $\Delta \sim 10^{15}$ cm. The luminosity in selected energy bands as well as the instantaneous and time integrated spectra of GRB 031203 are represented in Figs. \ref{031203_picco}-\ref{031203_spettro}. We here recall that such instantaneous spectra present a very clear hard-to-soft behavior, and that the corresponding time integrated spectrum is in very good agreement with the observed one. In Fig. \ref{urca123+GRB_full}C we compare the light curves of GRB 031203 in the $2$--$200$ keV energy band, of SN 2003lw in the optical band \citep{nomoto,pian06} and of URCA-3 observed by XMM-EPIC in the $0.2$--$10$ keV energy band \citep{watson} and by \textit{Chandra} in the $2$--$10$ keV energy band \citep{soderberg}.

\section{What are ``short'' and ``long'' GRBs?}

\begin{figure}
\includegraphics[width=\hsize,clip]{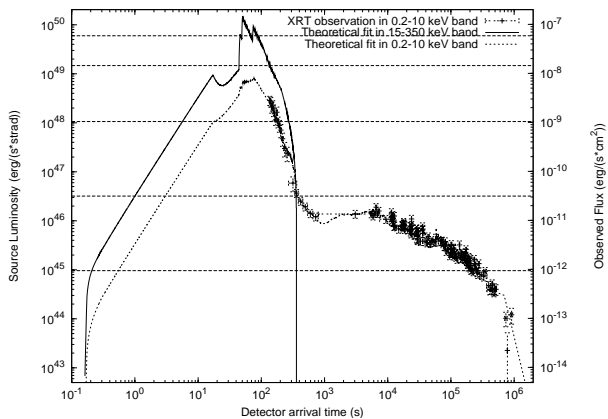}
\caption{Same as Fig. \ref{050315_global}. The horizontal dashed lines corresponds to different possible instrumental thresholds. It is clear that long GRB durations are just functions of the observational threshold.}
\label{global}
\end{figure}

The confirmation by \emph{Swift} of our prediction of the overall afterglow structure, and especially the coincidence of the ``prompt emission'' with the peak of the afterglow, opens a new problematic in the definition of the long GRBs. It is clear, in fact, that the identification of the ``prompt emission'' in the current GRB literature is not at all intrinsic to the phenomenon but is merely due to the threshold of the instruments used in the observations (e.g. BATSE in the $50$--$300$ keV energy range, or BeppoSAX GRBM in $40$--$700$ keV, or \emph{Swift} BAT in $15$--$350$ keV). As it is clear from Fig. \ref{global}, there is no natural way to identify in the source a special extension of the peak of the afterglow that is not the one purely defined by the experimental threshold. It is clear, therefore, that long GRBs, as defined till today, are just the peak of the afterglow and there is no way, as explained above, to define their ``prompt emission'' duration as a characteristic signature of the source. As the \emph{Swift} observations show, the duration of the long GRBs has to coincide with the duration of the entire afterglow. A Kouveliotou - Tavani plot of the long GRBs, done following our interpretation which is clearly supported by the recent \emph{Swift} data (see Fig. \ref{global}), will present enormous dispersion on the temporal axis.

\begin{figure}
\centering
\includegraphics[width=\hsize,clip]{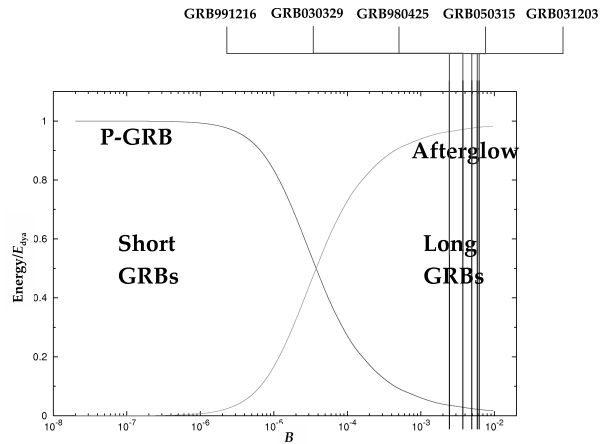}
\caption{The energy radiated in the P-GRB and in the afterglow, in units of the total energy of the dyadosphere ($E_{dya}$), are plotted as functions of the $B$ parameter. The values of the $B$ parameter computed in our theory for the sources GRB 991216, GRB 030329, GRB 980425, GRB 050315, GRB 031203 are also represented. It is very remarkable that they are all consistently smaller than the absolute upper limit $B < 10^{-2}$ found in \citet{rswx00}.}
\label{bcross}
\end{figure}

We recall that in our theory both ``short'' and ``long'' GRBs originate from the same process of black hole formation. The major difference between the two is the value of the baryon loading parameter $B$ (see Fig. \ref{bcross}). In the limit of small baryon loading, all the plasma energy is emitted at the transparency in the P-GRB, with negligible afterglow observed flux. For higher values of the baryon loading, the relative energy content of the P-GRB with respect to the afterglow diminishes \citep[see e.g.][and references therein]{rubr2}

\section{Conclusions}

\begin{table*}
\centering
\caption{a) see \citet{yk}; b) Mazzali, P., private communication at MG11 meeting in Berlin, July 2006; c) evaluated fitting the URCAs with a power law followed by an exponentially decaying part; d) evaluated assuming a mass of the neutron star $M=1.5 M_\odot$ and $T \sim 5$--$7$ keV in the source rest frame; e) see \citet{galama98,greiner,proch,mh}.}
{\footnotesize
\begin{tabular}{ccccccccccc}
\hline
GRB & $\begin{array}{c}E_{e^\pm}^{tot}\\ \mathrm{(erg)}\end{array}$ & $B$ & $\gamma_0$ & $\begin{array}{c}E_{SN}^{bolom}\\ \mathrm{(erg)^a}\end{array}$ & $\begin{array}{c}E_{SN}^{kin}\\ \mathrm{(erg)^b}\end{array}$ & $\begin{array}{c}E_{URCA}\\ \mathrm{(erg)^c}\end{array}$ & $\displaystyle\frac{E_{e^\pm}^{tot}}{E_{URCA}}$ & $\displaystyle\frac{E_{SN}^{kin}}{E_{URCA}}$ & $\begin{array}{c}R_{NS}\\ \mathrm{(km)^d}\end{array}$ & $z^e$ \\
\hline
980425 & $1.2\times 10^{48}$ & $7.7\times10^{-3}$ & $124$ & $2.3\times 10^{49}$ & $1.0\times 10^{52}$ & $3\times 10^{48}$ & $0.4$ & $1.7\times10^{4}$ & $ 8$ & $0.0085$\\
030329 & $2.1\times 10^{52}$ & $4.8\times10^{-3}$ & $206$ & $1.8\times 10^{49}$ & $8.0\times10^{51}$ & $3\times10^{49}$ & $6\times 10^{2}$ & $1.2\times10^{3}$ & $14$ & $0.1685$\\
031203 & $1.8\times 10^{50}$ & $7.4\times10^{-3}$ & $133$ & $3.1\times 10^{49}$ & $1.5\times10^{52}$ & $2\times10^{49}$ & $8.2$ & $3.0\times10^{3}$ & $20$ & $0.105$\\
060218 & $1.8\times 10^{50}$ & $1.0\times10^{-2}$ & $99$ & $9.2\times 10^{48}$ & $2.0\times10^{51}$ & $?$ & $?$ & $?$ & $?$ & $0.033$\\
\end{tabular}}
\label{tabella}
\end{table*}

In Tab. \ref{tabella} we summarize the representative parameters of the above four GRB-SN systems, including the very large kinetic energy observed in all SNe \citep{mazzaliVen}. Some general conclusions on these weak GRBs at low redshift, associated to SN Ib/c, can be established on the ground of our analysis:\\
{\bf 1)} From the detailed fit of their light curves, as well as their accurate spectral analysis, it follows that all the above GRB sources originate consistently from the formation of a black hole. This result extends to this low-energy GRB class at small cosmological redshift the applicability of our model, which now spans over a range of energy of six orders of magnitude from $10^{48}$ to $10^{54}$ ergs \citep{rubr,cospar02,030329,030329_la,031203,050315}. Distinctive of this class is the very high value of the baryon loading which in one case (GRB 060218) is very close to the maximum limit compatible with the dynamical stability of the adiabatic optically thick acceleration phase of the GRBs \citep{rswx00}. Correspondingly, the maximum Lorentz gamma factors are systematically smaller than the ones of the more energetic GRBs at large cosmological distances. This in turn implies the smoothness of the observed light curves in the so-called ``prompt phase''. The only exception to this is the case of GRB 030329.\\
{\bf 2)} The accurate fits of the GRBs allow us to infer also some general properties of the ISM. While the size of the clumps of the inhomogeneities is $\Delta \approx 10^{14}$ cm, the effective ISM average density is consistently smaller than in the case of more energetic GRBs: we have in fact $\langle n_{ism} \rangle$ in the range between $\sim 10^{-6}$ particle/$cm^3$ (GRB 060218) and $\sim 10^{-1}$ particle/$cm^3$ (GRB 031203), while only in the case of GRB 030329 it is $\sim 2$ particle/$cm^3$. We are also currently studying a characteristic trend in the variability of $\mathcal{R}$  during some specific bursts as well as the physical origin of the consistently smaller effective ISM density $\langle n_{ism} \rangle$ values observed in these sources \citep[see][]{060218}.\\
{\bf 3)} Still within their weakness these four GRB sources present a large variability in their total energy: a factor $10^4$ between GRB 980425 and GRB 030329. Remarkably, the SNe emission both in their very high kinetic energy and in their bolometric energy appear to be almost constant respectively $10^{52}$ erg and $10^{49}$ erg.\\
{\bf 4)} The URCAs present a remarkably steady behavior around a ``standard luminosity'' and a typical temporal evolution. If, indeed, their connection to GRBs will be excluded by future Swift observations, then we should consider the possibility that the URCAs are related to the SNe: either to dissipative processes in the SN ejecta, or to the formation of a neutron star in the SN explosion \citep{mg10}. As an example, we have given an estimate of the corresponding neutron star radius for URCA-1, URCA-2 and URCA-3 (see Tab. \ref{tabella}). It has been pointed out \citep[see e.g.][]{pian00} the different spectral properties between the GRBs and the URCAs. It would be interesting to compare and contrast the spectra of all URCAs in order to evidence the expected analogies among them. Observations of the URCA sources on time scales of $0.1$--$10$ seconds would be highly desirable.

\begin{figure}[t]
\includegraphics[width=\hsize,clip]{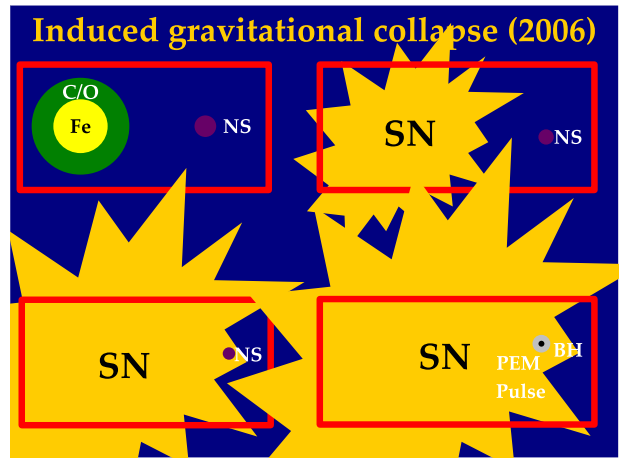}
\caption{A possible process of gravitational collapse to a black hole ``induced'' by the Ib/c SN on a companion neutron star in a close binary system. Details in \citet{mg11}.}
\label{IndColl06}
\end{figure}

The majority of models in the current literature attribute the GRB and SN origin to a single star system \citep[``collapsar'', see e.g.][]{wb06}. We proposed, instead, that such GRBs associated to SNe originate in a binary system. The GRB originates from the collapse to a black hole, while the SN originates from the collapse of a companion star evolved out of the main sequence. We expect that the temporal coincidence of the effects is due to the induced gravitational collapse phenomenon. In this respect, there is the possibility that some of the above features may be accounted for by a process of gravitational collapse to a black hole ``induced'' by the Ib/c SN on a companion neutron star in a close binary system. The weakness in the energetics of GRB 980425 and GRB 031203, and the sizes of their dyadospheres, evidence that they originate in the formation of the smallest possible black hole, just over the critical mass of the neutron star \citep[see Fig. \ref{IndColl06} and][]{mg11}.

\end{document}